# High carrier mobility in single-crystal $PtSe_2$ grown by molecular beam epitaxy on ZnO(0001)

Frédéric Bonell[1], Alain Marty[1], Céline Vergnaud[1], Vincent Consonni[2], Hanako Okuno[3], Abdelkarim Ouerghi[4], Hervé Boukari[5], and Matthieu Jamet[1]

[1] Univ. Grenoble Alpes, CNRS, CEA, Spintec, 38000 Grenoble, France
[2] Univ. Grenoble Alpes, CNRS, Grenoble INP, LMGP, 38000 Grenoble, France
[3] Univ. Grenoble Alpes, CEA, IRIG-MEM, 38000 Grenoble, France
[4] Univ. Paris-Saclay, CNRS, Centre de Nanosciences et de Nanotechnologies, 91120, Palaiseau, France
[5] Univ. Grenoble Alpes, CNRS, Grenoble INP, Institut Néel, 38000 Grenoble, France

E-mail: frederic.bonell@cea.fr

**Abstract**

$PtSe_2$ is attracting considerable attention as a high mobility two-dimensional material with envisionned applications in microelectronics, photodetection and spintronics. The growth of high quality $PtSe_2$ on insulating substrates with wafer-scale uniformity is a prerequisite for electronic transport investigations and practical use in devices. Here, we report the growth of highly oriented few-layers $PtSe_2$ on ZnO(0001) by molecular beam epitaxy. The crystalline structure of the films is characterized with electron and X-ray diffraction, atomic force microscopy and transmission electron microscopy. The comparison with $PtSe_2$ layers grown on graphene, sapphire, mica, $SiO_2$ and Pt(111) shows that among insulating substrates, ZnO(0001) yields films of superior structural quality. Hall measurements performed on epitaxial $ZnO/PtSe_2$ with 5 monolayers of $PtSe_2$ show a clear semiconducting behaviour and a high mobility in excess of 200 $cm^2V^{-1}s^{-1}$ at room temperature and up to 447 $cm^2V^{-1}s^{-1}$ at low temperature.

## 1. Introduction

Transition metal dichalcogenides (TMDs) form a vast family of two-dimensional (2D) materials with outstanding properties arising from the strong electronic confinement in their van der Waals (vdW)-bonded layers, and making them extremely attractive for applications in nanoelectronics, optoelectronics, spintronics and gas sensing [1-3].

Among the most studied TMDs, group-VI compounds such as $MoS_2$ or $WSe_2$ adopt the hexagonal 2H structure and are semiconducting, with a 1-2 eV band gap energy and a moderate carrier mobility reaching few 100 $cm^2V^{-1}s^{-1}$ [4]. More recently, group-X Pd- and Pt-based compounds have emerged as interesting TMDs with distinct properties [5,6]. Their larger *d*-band filling favours octahedral coordination of the TM element (trigonal 1T structure) and an ensuing bulk semimetallic character. 1T-TMDs are characterized by a stronger interlayer hybridization than in 2H-TMDs, which is responsible for a marked thickness dependence of their electronic properties.

Of special interest is bulk $1T-PtSe_2$, an air-stable type-II Dirac semimetal [7] that undergoes an unusual metal-to-semiconductor transition when thinned down to a few layers [8-11]. Its band gap is indirect and reaches an energy of ~1.2 eV for a monolayer (ML). This tunable band gap offers interesting opportunities for broadband mid-infrared photodetection [12] and versatile field-effect transistors. Importantly, $PtSe_2$ is expected to host high mobility carriers, with a theoretical phonon-limited mobility larger than 3000 $cm^2V^{-1}s^{-1}$ at room temperature [13,14] (compared to 400 $cm^2V^{-1}s^{-1}$ in $MoS_2$ [4]). $PtSe_2$ additionally exhibits superior photocatalytic properties [8] and remarkable magnetic characteristics. Long-range spin ordering induced by native defects has been reported in mono- and bilayers [15]. $PtSe_2$ also displays a local Rashba effect resulting in spin-layer locking. Helical spin textures have experimentally been observed in bulk and monolayers [16,17], and may be exploited in electrostatically-controlled spintronics. Furthermore, it has recently been shown that the inversion symmetry of the 1T structure can be lifted by converting a single-ML $PtSe_2$ into a Janus SPtSe compound [18], enabling a strong Rashba effect.

Single- and few-layers-thick $PtSe_2$ have been prepared by a variety of methods, including mechanical exfoliation [9,10,14,15], selenization or thermally-assisted conversion (TAC) [8,19- 21], chemical vapor deposition (CVD) [11,22] and molecular beam epitaxy (MBE) [17]. However, the fabrication of films that are uniform at the wafer scale and exhibit a well-controlled thickness and high crystallinity remains highly challenging. Crystal imperfections contribute to limiting the carrier mobility well below the expectations for pristine layers. Apart from few experimental reports of carrier mobilities higher than 200 $cm^2V^{-1}s^{-1}$ [14,20], the measured values are generally of the order of 10 $cm^2V^{-1}s^{-1}$ [9-11].

$PtSe_2$ layers with the presumably highest crystal quality have been formed by direct selenization of a Pt(111) single crystal [8]. The reaction is self-limited and yields a single-crystalline $PtSe_2$ monolayer on top of Pt(111). The bonding with the Pt(111) buffer turns to be strong [18] and its exact nature has not been elucidated yet [23]. There have been several attempts to adapt this process to thin Pt layers deposited on insulating substrates, which are suitable for electrical measurements. Selenization of a 3 ML-thick Pt film grown by MBE on sapphire resulted in the incomplete conversion into $PtSe_2$ [21]. TAC was also performed in a CVD reactor on 0.75-10-nm-thick Pt sputtered onto $Si/SiO_2$ [19,20]. In that case, the higher Se supply rate and higher process temperature resulted in the full conversion of the films but yielded polycrystalline $PtSe_2$.

Recently, high quality growth of $PtSe_2$ (1-22 ML) on SiC(0001)/graphene by MBE has been reported [17]. The MBE technique presents several advantages, such as the ultra-high vacuum (UHV) environment, the high purity of the evaporated materials, the large scale uniformity, the precise control of the thickness and the possibility to grow heterostructures with sharp interfaces. The epitaxy of layered materials down to single monolayers has been shown for a variety of TMDs [17,24-30]. On the one hand, substrates with a vdW termination like graphene or mica are generally preferred. Their inertness makes it possible to grow continuous monolayers with preserved vdW properties. On the other hand, the weak substrate-layer interaction may result in a large in-plane misorientation of the crystal grains (i.e. mosaicity). Quasi-vdW epitaxy on non-vdW surfaces can possibly promote a better crystal alignment, as shown for instance with $BaF_2(111)/Bi_2Te_3$ [31] or $InAs(111)/ZrTe_2$ [27].

The MBE growth of $PtSe_2$ on insulators, as well as quasi-vdW epitaxy of $PtSe_2$ have not been investigated so far. In this article, we show that highly oriented $PtSe_2$ can be grown on ZnO(0001) single crystals. ZnO is an eco-friendly material composed of abundant elements and a wide bandgap semiconductor with piezoelectric properties and potential applications in optoelectronics [32]. We also provide a comparative study of epitaxial $PtSe_2$ films grown on SiC/graphene and on several insulating substrates, including mica and $Al_2O_3$(0001) (sapphire). The crystalline quality is assessed with reflection high energy electron diffraction (RHEED) and grazing incidence X-ray diffraction (GI-XRD), and compared to the one of $PtSe_2$ layers formed by selenization of Pt(111) and by selenization of Pt sputtered on $Si/SiO_2$. Additional atomic force microscopy (AFM) and transmission electron microscopy (TEM) characterization is provided. We show that among the investigated insulating substrates, ZnO(0001) yields the $PtSe_2$ layers of the highest structural quality. Furthermore, Hall measurements performed on epitaxial $ZnO/PtSe_2$ show a clear n-type semiconducting behaviour and a mobility of 447 $cm^2V^{-1}s^{-1}$ at low temperature.

**2. Methods**

The films were grown under UHV ($2\times10^{-9}$ mbar base pressure) in an MBE chamber equipped with cryo-panels, a RHEED setup and a sample manipulator with radiation/electron bombardment heater. Pt and Se fluxes were respectively supplied by electron-beam evaporation and standard effusion with a Knudsen cell. The MBE reactor is connected under UHV to a magnetron sputtering chamber with a Pt target and a retractable quartz crystal microbalance.

The substrates were conditionned as follows. $Si/SiO_2$ and sapphire were first cleaned with acetone and isopropanol in an ultrasonic bath. Sapphire was then annealed for 1 h at 1000°C in air. $Si/SiO_2$, sapphire and 6H-SiC(0001)/graphene were annealed for 30 min at 800°C in UHV prior to the growth. Fresh mica surfaces (*Ted Pella Inc.*) were obtained by mechanical exfoliation followed by UHV annealing for 5 min at 800°C. ZnO(0001) substrates (*CrysTec GmbH*) were initially etched for 30 s with an HCl 1.8% solution and then rinsed with deionized water. This chemical treatment was followed by a thermal annealing for 1 h at 900°C under $O_2$ atmosphere in a tubular furnace, and by a subsequent thermal annealing in UHV for 5 min at ~900°C. The Zn-terminated face of ZnO was preferred for its simpler preparation and to avoid the direct proximity between oxygen and $PtSe_2$.

For the growth of $PtSe_2$ by coevaporation, the Pt deposition rate was set to 0.3-0.5 Å/min and monitored in real time by a quartz crystal microbalance. The beam equivalent pressure of Se at the sample position was measured with a retractable ionization gauge and set to $1\times10^{-6}$ mbar, which corresponds to a Se:Pt flux ratio of about 10~20. The substrate temperature was varied in the range of 300-600°C. In the following, we focus on the growth temperature of 490°C that yields the best crystal quality, as inferred from RHEED observations. All PtSe$_2$ layers were annealed at 800°C for 10 min under Se flux after the growth in order to improve their crystalline quality.

GI-XRD measurements were performed with a SmartLab Rigaku diffractometer. The source was a rotating anode beam tube (Cu K$_\alpha$ = 1.54 Å) operating at 45 kV and 200 mA. The diffractometer was equipped with a parabolic multilayer mirror and in-plane collimators of 0.5° on both the source and detector sides, defining the angular resolution. A K$_\beta$ filter on the detector side was used to eliminate parasitic radiations.

## 3. Results and discussion

### 3.1. ZnO(0001)/PtSe$_2$ growth and characterization

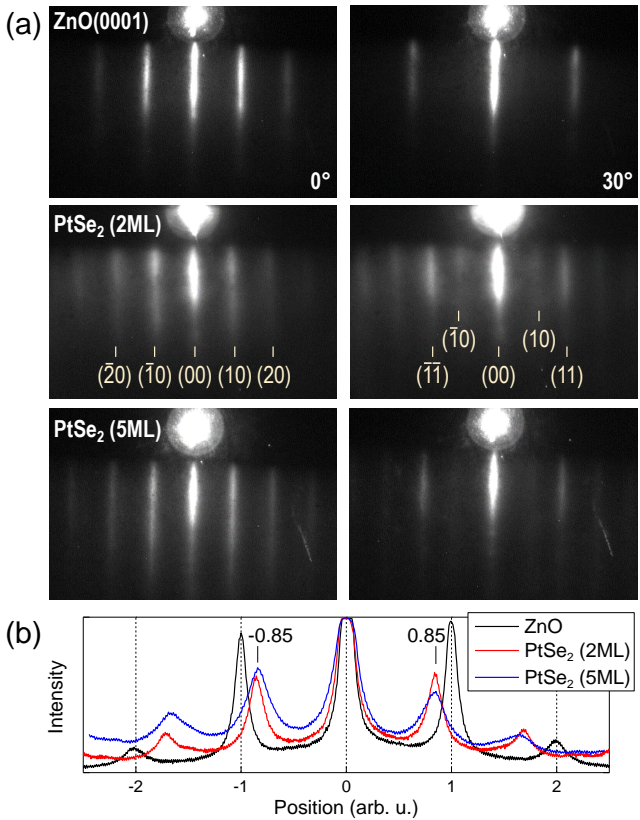

*Figure 1. (a) RHEED patterns of ZnO(0001) and ZnO/PtSe$_2$ in two different azimuths. (b) Intensity profile of the (h0) patterns (0°), showing the thickness-independent relaxation of PtSe$_2$ and the lattice misfit with ZnO(0001).*

Fig. 1 shows the RHEED patterns of the ZnO(0001) surface and those of 2 ML and 5 ML-thick PtSe$_2$ layers after post-growth annealing. In all cases, the diffraction patterns are anisotropic and display a 6-fold symmetry. The epitaxy relationship is found to be ZnO(0001)[100]/PtSe$_2$(0001)[100]. The faint (h0) rods visible on the (hh) patterns of PtSe$_2$, especially in the 2 ML-thick film, reveal the presence of twisted grains. The in-plane surface lattice parameter deduced from the distance between the rods is close to the bulk value $a_{PtSe_2}$ = 0.373 nm for both 2 ML and 5 ML films (Fig. 1b). This is 15% larger than the lattice parameter of ZnO ($a_{ZnO}$ = 0.325 nm). The fact that the films are fully relaxed is typical of (quasi-)vdW epitaxy and therefore suggests a weak interfacial interaction between ZnO and PtSe$_2$. However, we note that during the initial stages of PtSe$_2$ deposition the RHEED patterns systematically reveal a highly disordered first ML. Diffraction rods appear during the growth of the second ML and become more intense with further deposition. Intensity oscillations were not observed.

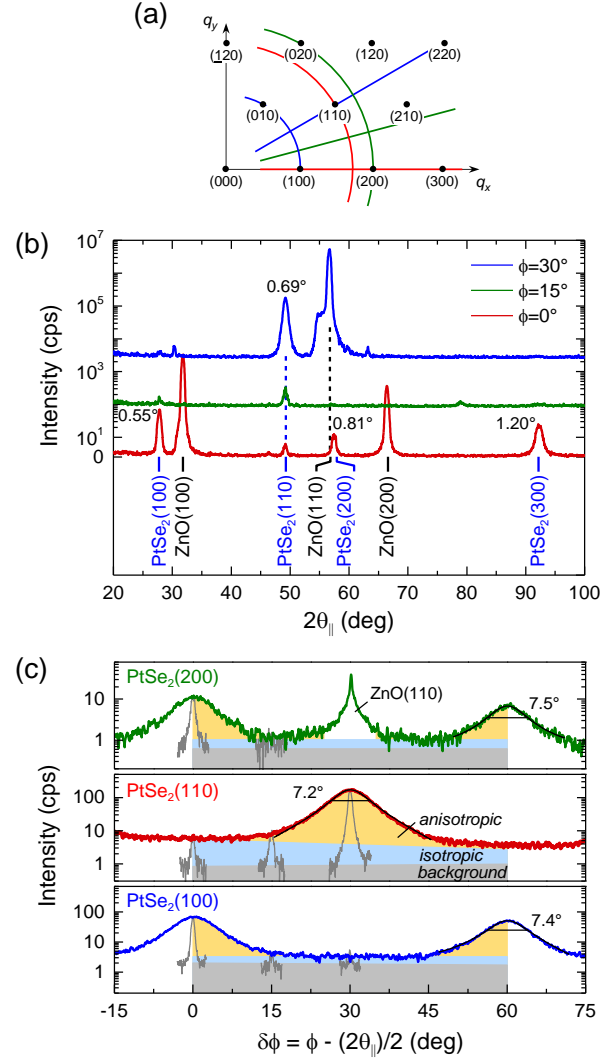

*Figure 2. GI-XRD on ZnO(0001)/PtSe$_2$ (5ML). (a) Reciprocal space trajectories of the displayed scans. (b) Radial scans for three in-plane orientations. The numbers above individual peaks are their FWHM. (c) Azimuthal scans across the (100), (110) and (200) PtSe$_2$ reflections. The partial radial scans displayed at 0°, 15° and 30° were used to evaluate the background and isotropic/anisotropic contributions to the signal (filled areas).*

The GI-XRD results on the 5 ML-thick film are presented in Fig. 2. The Bragg reflections in the radial scans (Fig. 2a) correspond to an average lattice parameter $\langle a_{PtSe2}\rangle = 0.370$ nm, 0.65% smaller than the bulk value. The azimuthal scans (Fig. 2b) confirm the highly anisotropic angular distribution of the crystallites. The full width at half maximum (FWHM) of the peaks, which reflects the in-plane rotational misorientation of the grains, amounts to $\Delta\Phi = 7.4\pm0.2°$ [when averaging over (100), (110) and (200) reflections]. The most intense (110) peak is visible in all radial scans. Consistently with the above discussion of the RHEED patterns, this indicates the presence of grains with random in-plane orientation. Their relative volume is estimated from the area enclosed between the base line of the azimuthal scans and the background level determined from the base line of radial scans. This analysis is explicited in Fig. 2c, where the relevant portions of the radial scans at 0°, 15° and 30° are delineated. We find an isotropic contribution in the diffracted intensity of 15% for this film. An alternative analysis yields 86% (resp. 68%) of the crystallites oriented within ±10° (resp. ±5°) from the main epitaxial direction.

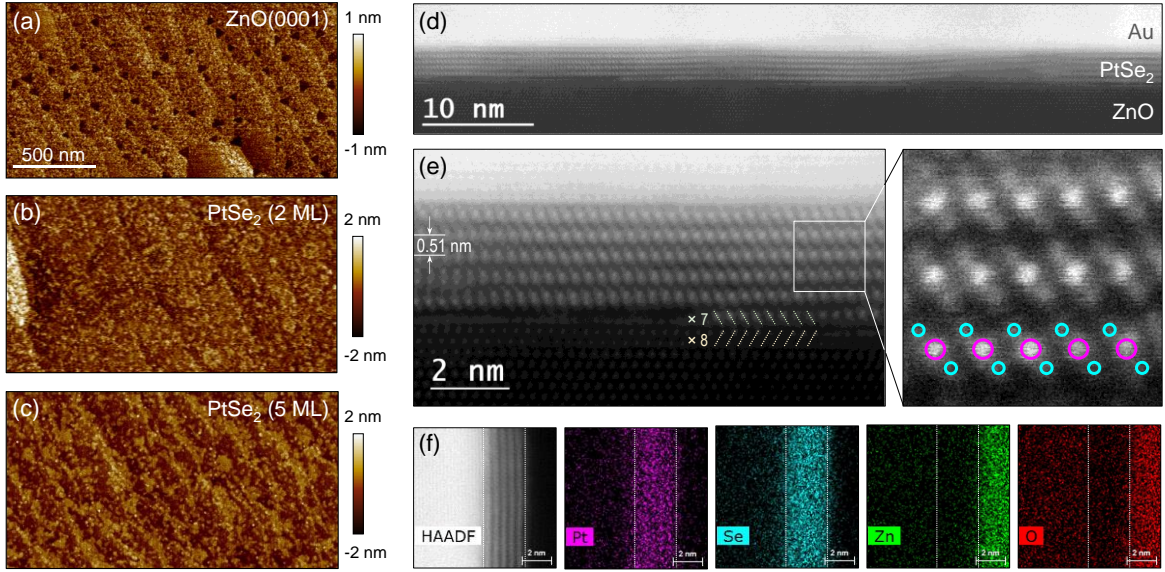

*Figure 3. (a-c) Surface morphology measured by AFM (same scale on all images). (a) ZnO(0001) substrate after chemical treatment and thermal annealing under $O_2$ atmosphere. (b) 2-ML-thick $PtSe_2$(0001) grown on ZnO(0001). (c) 5-ML-thick $PtSe_2$(0001). (d-f) STEM cross-sections of ZnO/$PtSe_2$ (5 ML)/Au. In (e), the 7:8 correspondence between $PtSe_2$ and ZnO interatomic spacing is highlighted. (f) EDX analysis showing the chemical uniformity of $PtSe_2$ and the absence of intermixing at the ZnO/$PtSe_2$ interface.*

Fig. 3a-c show an AFM image of the ZnO(0001) surface after the thermal annealing under $O_2$ atmosphere together with images of the 2 ML and 5 ML-thick $PtSe_2$ films. The ZnO(0001) surface displays terraces with a 130 nm average width, which are mostly delimited by monolayer steps with a 0.5 nm height. In addition, 0.5-nm-deep triangular pits resulting from the HCl etching are distributed along the atomic steps. The root-mean-squared (rms) roughness of the ZnO(0001) surface (Fig. 3a) is 0.28 nm. The terraces of the substrate are preserved after deposition of $PtSe_2$ (Fig. 3b) The 2-ML-thick film uniformly covers the surface but presents a grainy morphology, with a 0.53 nm rms roughness. The 5-ML-thick film (Fig. 3c) presents more extended flat areas and accordingly, a reduced rms roughness of 0.42 nm.

We additionally examined the atomic structure of $PtSe_2$ with scanning transmission electron microscopy in the high angle annular dark field mode (HAADF-STEM). Cross-section images shown in Fig. 3d,e confirm the high crystallinity of $PtSe_2$ and reveal a sharp ZnO/$PtSe_2$ interface. In Fig. 3f, chemical analysis by energy dispersive X-ray spectroscopy (EDX) demonstrates the absence of atomic interdiffusion at the interfaces.

### 3.2. Comparison with $PtSe_2$ grown on other substrates

In the following, we compare the structural quality of $PtSe_2$ grown on ZnO(0001) single crystals and on other substrates. We first consider two well-studied cases: Pt(111)/$PtSe_2$ and $SiO_2$/$PtSe_2$. Both were obtained by room temperature selenization of Pt followed by annealing at 800°C under exposure to a Se flux. In the case of Pt(111)/$PtSe_2$, the single-crystalline Pt layer (30 nm) was grown by MBE on sapphire. For $SiO_2$/$PtSe_2$, the Pt layer was deposited by DC sputtering in a chamber connected to the MBE reactor. The Pt thickness was set to 0.23 nm in order to form a 2-ML-thick $PtSe_2$ film after selenization.

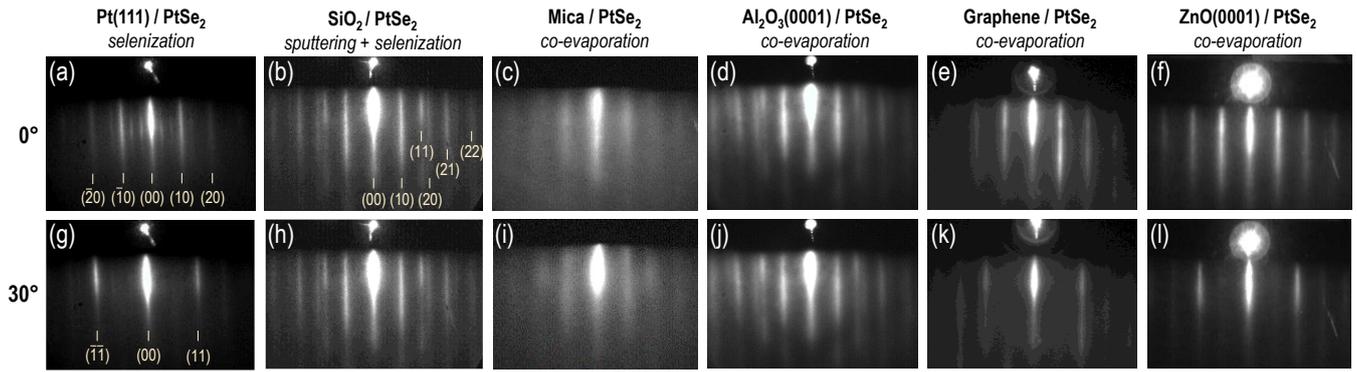

*Figure 4. RHEED patterns PtSe$_2$ grown on various substrates, in two different azimuths (0° and 30°). The highly anisotropic diffraction patterns of Pt(111)/PtSe$_2$, graphene/PtSe$_2$ and ZnO(0001)/PtSe$_2$ demonstrate epitaxial growth. In contrast, the isotropic patterns of SiO$_2$/PtSe$_2$, Al$_2$O$_3$(0001)/PtSe$_2$ and mica/PtSe$_2$ reveal a mostly random in-plane orientation of the grains. The PtSe$_2$ thickness is 1 ML on Pt(111), 2 ML on SiO$_2$ and 5 ML on the other substrates.*

PtSe$_2$ on Pt(111) is single-crystalline, as evidenced by the anisotropic RHEED patterns in Fig. 4a,g. On the other hand, the amorphous SiO$_2$ cannot promote in-plane orientation of PtSe$_2$ and therefore, results in a fully polycrystalline film and isotropic RHEED patterns where all reflections are visible in a single diffraction pattern (Fig. 4b,h). We additionally investigated PtSe$_2$ grown on mica, sapphire and SiC/graphene by co-evaporation. On mica and sapphire, our best growth conditions were found similar to those used for the growth on ZnO(0001): deposition at a substrate temperature of ~490°C followed by annealing at 800°C under Se flux. On SiC/graphene, the growth was performed at a lower substrate temperature of 300°C. The corresponding RHEED patterns for 5-ML-thick films are displayed in Fig. 4c-e,i-k. The diffraction patterns of both mica/PtSe$_2$ and sapphire/PtSe$_2$ are mainly isotropic. In contrast, those of graphene/PtSe$_2$ are largely anisotropic, which indicates a single-crystalline PtSe$_2$ layer, in line with previous reports [17].

Quantitative assessment of the mosaicity and crystal anisotropy was performed by the analysis of GI-XRD azimuthal scans. Fig. 5a compares the (110) scans for all samples, where the isotropic/anisotropic components of the signal are highlighted. In the case of Pt(111)/PtSe$_2$ and graphene/PtSe$_2$, no isotropic contribution is evidenced within the experimental accuracy, whereas the SiO$_2$/PtSe$_2$ film is fully isotropic. The satellite peaks at ±20° from the Pt/PtSe$_2$(110) reflection originate from the 4:3 exact coincidence between the Pt and PtSe$_2$ lattices [8,18]. Correspondingly, the (3×3) PtSe$_2$ superstructure results in additional diffraction rods between (h0) streaks in the RHEED patterns of Fig. 4a. Consistently with the RHEED observations, the highest crystal orientation for PtSe$_2$ grown on insulating substrates is found on ZnO(0001). The mosaicity turns to be slighly lower than in graphene/PtSe$_2$. In contrast, the azimuthal scans of PtSe$_2$ on both mica and sapphire show a dominant isotropic component. The ZnO/PtSe$_2$(110) and graphene/PtSe$_2$(110) scans are well fitted with a single Lorentzian line shape, whereas mica/PtSe$_2$(110) and sapphire/PtSe$_2$(110) scans turn to be best fitted with a sum of three Lorentzian peaks centered at 30° and 30°±ΔΦ, which suggests preferential twist angles from the main epitaxial orientation, at ±15° on mica and ±30° on sapphire.

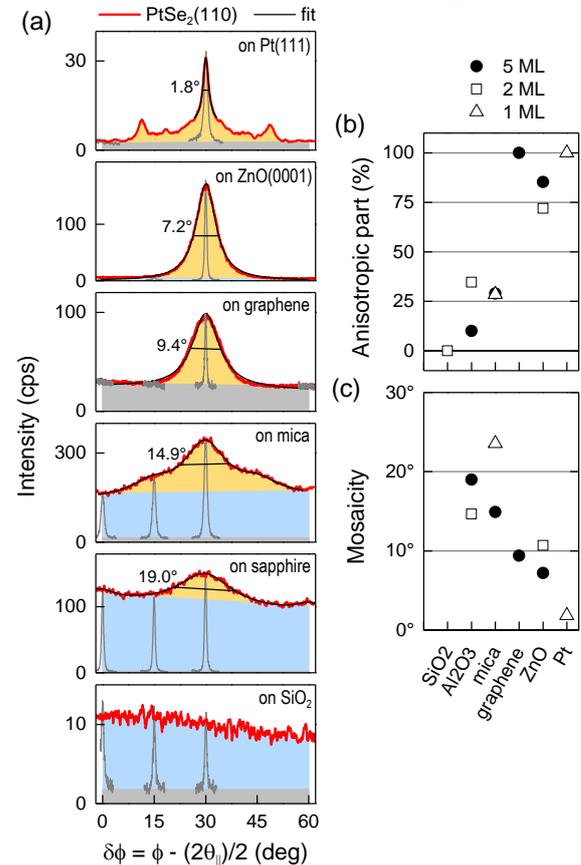

*Figure 5. (a) Azimuthal GI-XRD scans across the PtSe$_2$(110) reflection for PtSe$_2$ films grown on various substrates [1 ML on Pt(111), 2 ML on SiO$_2$, 5 ML on others]. The partial radial scans displayed at 0°, 15° and 30° are used to evaluate the background and isotropic/anisotropic contributions to the signal (filled areas). (b) Anisotropic contribution to the diffracted intensity. (c) FWHM of the (110) peak centered at 30°, determined by fitting with a Lorentzian line shape.*

The anisotropic contribution to the signal and the FWHM of the (110) peak are reported in Fig. 5b,c. We also added to these figures the results obtained with thinner films. 2-ML-thick films on ZnO(0001) are slighlty less oriented than thicker ones but mostly remain epitaxial. For mica/PtSe$_2$ (1 ML), the anisotropic contribution is found equal to the one in the 5-ML-thick film. The larger FWHM might be due to the fact that the $\pm15°$ shoulders on the (110) peak could not be resolved for this film. In the case of Al$_2$O$_3$/PtSe$_2$ (2 ML), an annealing was performed after the growth of each ML. This procedure apparently resulted in a slighlty higher orientation of the grains, although far from the epitaxial alignement of ZnO/PtSe$_2$.

### 3.3. Transport in ZnO(0001)/PtSe$_2$

Transport measurements were performed on a 5-ML-thick ZnO/PtSe$_2$ layer capped in situ with 2.5 nm of Al that was later naturally oxidized in air. 100×20 µm$^2$ Hall crosses were patterned by laser lithography and Ar etching (Fig. 6a). The evolution of the resistivity as a function of temperature is presented in Fig. 6b. The film exhibits a metallic behaviour from room temperature down to 190 K, indicative of the activation of phonon scattering. Below 190 K, the sheet resistance $R_□$ sharply increases when lowering the temperature. The resistivity $\rho$ in the range 50 K-150 K is well fitted by an activation model corresponding to a band conduction mechanism, namely, $\rho = \rho_0 \exp(\Delta E_a/k_B T)$ with $k_B$ the Boltzmann constant and an activation energy $\Delta E_a = 22$ meV. Below 50 K, the resistivity clearly deviates from this model. In the range of 25 K-100 K, $\rho$ is better fitted with a 2D ($d = 2$) variable range hopping (VRH) mechanism, $\rho = \rho_0 \exp[(T_0/T)^{1/(d+1)}]$ with $T_0 = 9.8\times10^4$ K. Above 100 K, this model fails to describe correctly the measured temperature dependence. Note that VRH with $d = 3$ and $T_0 = 3.9\times10^6$ K (not shown) fits equally well the experimental data.

The Hall resistance curves in Fig. 6c reveal a negative slope $R_H$ that indicates electron transport. The longitudinal magnetoresistance (not shown) remains negligible in the whole temperature range. From these electrical data, we extract the carrier density $n = 1/(eR_H t)$ and the mobility $\mu = R_H/R_□$ that are reported in Fig. 6d,e, respectively. $e$ is the elementary charge and $t = 2.5$ nm the thickness. The results are shown for two devices fabricated on two films grown separately in identical conditions. The carrier mobility in device 1 (resp. device 2) reaches 216 cm$^2$V$^{-1}$s$^{-1}$ (resp. 207 cm$^2$V$^{-1}$s$^{-1}$) at room temperature and 432 cm$^2$V$^{-1}$s$^{-1}$ (resp. 447 cm$^2$V$^{-1}$s$^{-1}$) at 50 K. We attribute these remarkably high carrier mobilities to the high crystalline quality of epitaxial ZnO/PtSe$_2$.

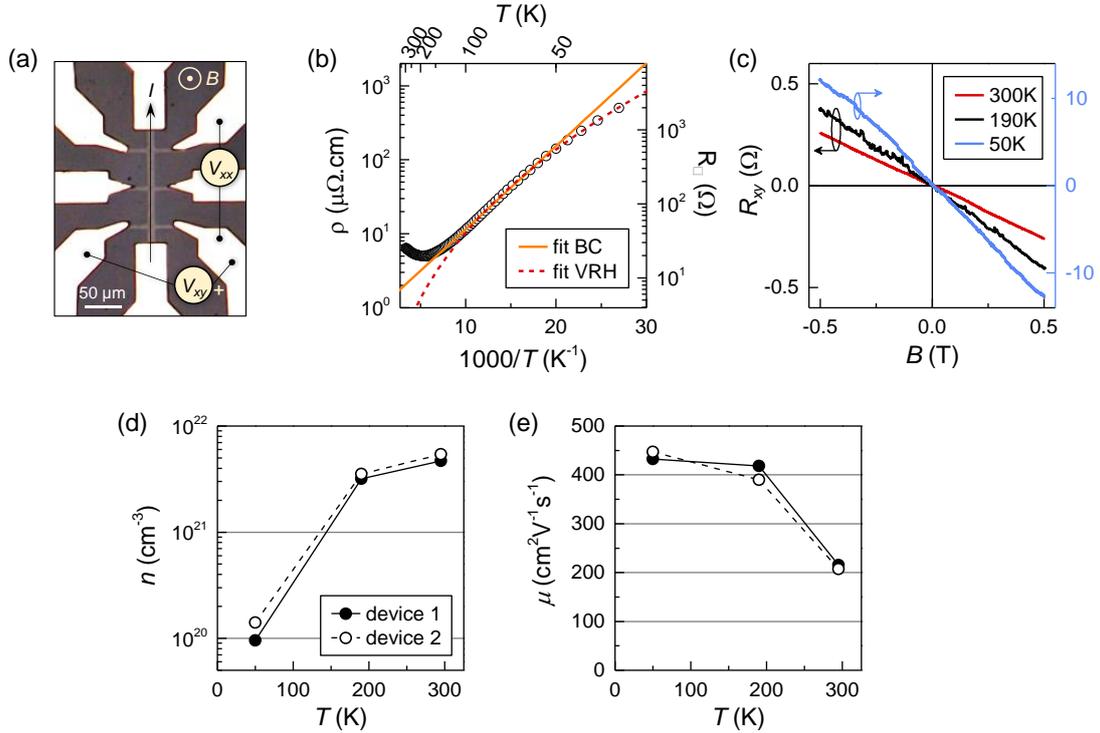

*Figure 6. Resistivity and Hall measurements on ZnO/PtSe$_2$ (5 ML)/AlO$_x$. (a) Picture of a Hall device. I is the applied charge current, B the out-of-plane magnetic field, V$_{xx}$ and V$_{xy}$ the measured longitudinal and Hall voltages, respectively. (b) Arrhenius plot of the resistivity (left axis) and sheet resistance (right axis), fitted with band conduction (BC) and 2D variable range hopping (VRH) models. (c) Transverse resistance at various temperatures. (d) Electron density n and (e) mobility µ versus temperature extracted from resistivity and Hall measurements on two different devices.*

## 3.4. Discussion

The determination of the dominant conduction mechanism in the range of 50-100 K turns to be difficult on the basis of the only fit analysis. However, the high carrier mobility values are little compatible with the VRH mechanism. The latter refers to sequential hopping between localized states by inelastic tunnelling, which indicates disorder and localized defects. 2D VRH has been reported for monolayers of other semiconducting TMDs such as $MoS_2$ [33,34], $MoSe_2$ [28,35,36] and $MoTe_2$ [35], and systematically associated with low carrier mobilities in the range 1-10 $cm^2V^{-1}s^{-1}$. Therefore, the high carrier mobility, together with its weak sensitivity to the temperature ranging from 50 K to 200 K, suggests that the conduction above 50 K is largely dominated by band conduction and that VRH becomes sizable only below 50 K. Our determined activation energy may indicate a donor level at $2\Delta E_a \approx 45$ meV below the conduction band minimum, which sets a lower bound for the bandgap of 5-ML-thick $PtSe_2$. Band structure calculations predicted a bandgap of ~1.2 eV for a $PtSe_2$ monolayer, strongly reduced to 0.2 eV in bilayers [8,9]. We are not aware of theoretical values for thicker films. Experimentally, a bandgap of 2.2 eV (upper bound) was reported in 2.5-nm-thick flakes [10]. Our estimate is in line with those values.

The onset of VRH conduction at low temperature indicates the presence of localized states due to bulk or interfacial defects. Obviously, the $PtSe_2$ layers present several crystal imperfections, such as a finite mosaicity, surface steps (Fig. 3b,c) and lattice distortions (Fig. 3d). The carrier mobility, although reaching unprecedented levels, remains also below the theoretical expectations. We actually believe that room remains for improving the crystallinity of $PtSe_2$ films by optimizing the preparation of ZnO(0001). We noticed that the crystalline orientation of the films is highly sensitive to it. The growth on substrates prepared without either chemical etching or thermal annealing under oxygen atmosphere results in the formation of films with larger mosaicity (~12°) and a dominant isotropic component (~70%) as deduced from GI-XRD measurements. Even after chemical and thermal treatment, the ZnO(0001) surface remains defective, as evidenced by AFM imaging through the presence of triangular pits for instance (Fig. 3). An optimized protocol for ZnO preparation, possibly involving annealing at higher temperature or the use of an etchant weaker than HCl, is likely to result in improved semiconducting characteristics and higher carrier mobilities.

Finally, the mechanism that drives the epitaxy of $PtSe_2$ on ZnO and that accounts for the differences between the investigated substrates remains unclear at this stage. The lattice mismatch is large in all cases and seems to play a minor role here (+15% with ZnO(0001), -22% with sapphire, -28% with mica). The observed ZnO(0001)[100]/$PtSe_2$(0001)[100] relationship approximately matches the incommensurate 7:8 relationship (+0.4% mismatch), as highlighted in Fig. 3e. It is interesting to note that the alternative ZnO(0001)[100]/$PtSe_2$(0001)[110] relationship (hexagonal $PtSe_2$ lattice twisted by 30° relative to the one of ZnO) is not favoured in spite of the +0.7% misfit. Further investigation of the growth kinetics and grain formation during the initial stages of the growth is required to clarify this point.

## 4. Conclusion

In summary, we achieved highly oriented epitaxial growth of few-layers $PtSe_2$ on ZnO(0001) by molecular beam epitaxy and demonstrated the superior structural quality of the films compared to those grown on graphene, sapphire, mica or $SiO_2$. Hall measurements on epitaxial ZnO/$PtSe_2$ (5 ML) revealed clear semiconducting characteristics, with a bandgap larger than 45 meV and a remarkably high carrier mobility in excess of 200 $cm^2V^{-1}s^{-1}$ at room temperature and up to 447 $cm^2V^{-1}s^{-1}$ at low temperature. These results confirm the high potential of $PtSe_2$ for 2D-materials-based microelectronics.

## Acknowledgements

The authors acknowledge financial support from the European Union Horizon 2020 FETOPEN program under project NANOPOLY (Grant Agreement No. 829061), from the French ANR under project ELMAX (Grant No. ANR-20-CE24-0015) and from the King Abdullah University of Science and Technology under Grant No. ORS-2018-CRG7-3717. The authors also acknowledge the LANEF framework (ANR-10-LABX-51-01) for its support with mutualized infrastructure.